\documentclass[10pt,twocolumn,english,prd,superscriptaddress,nofootinbib,preprintnumbers,showpacs,floatfix]{revtex4-2}

\usepackage[utf8]{inputenc}
\usepackage{amsmath}
\usepackage{amsfonts}
\usepackage{amssymb}
\usepackage{graphicx} 
\usepackage[caption=false]{subfig}
\graphicspath{ {figures/} }
\usepackage[usenames,dvipsnames]{xcolor}
\usepackage{hyperref}   
\hypersetup{
	colorlinks=true, 
	citecolor=MidnightBlue,
	linkcolor=MidnightBlue,
	urlcolor=Cyan}
\usepackage{tikz}
\usepackage{verbatim} 
\usepackage{soul} 
\definecolor{purple}{rgb}{1,0,1}
\definecolor{lime}{HTML}{A6CE39} 

\newcommand{\orcidicon}{%
	\begin{tikzpicture}
		\draw[lime, fill=lime] (0,0) 
		circle [radius=0.16] 
		node[white] {{\fontfamily{qag}\selectfont \tiny ID}};
		\draw[white, fill=white] (-0.0625,0.095) 
		circle [radius=0.007];
	\end{tikzpicture}	\hspace{-2mm}
}

\newcommand\orcidClaudio{{\href{https://orcid.org/0000-0001-6022-459X}{\orcidicon}}}

\newcommand\orcidMargarida{{\href{https://orcid.org/0000-0003-2231-4978}{\orcidicon}}}

\newcommand\orcidFrancisco{{\href{https://orcid.org/0000-0002-9388-8373}{\orcidicon}}}

\newcommand\orcidLuis{{\href{https://orcid.org/0009-0009-4322-6484}{\orcidicon}}}
\begin{document}
	
	\title{Black hole shadows in nonminimally coupled Weyl connection gravity}
	
	\author{ Cl\'{a}udio Gomes\orcidClaudio\!\!}
	\email{claudio.fv.gomes@uac.pt}
	\affiliation{OKEANOS–Instituto de Investigaç\~{a}o em Ci\^{e}ncias do Mar, Universidade dos Açores,
		Rua Prof. Doutor Frederico Machado, 4, 9900-140 Horta, Portugal}
	\affiliation{Centro de F\'{i}sica das Universidades do Minho e do Porto,
		Rua do Campo Alegre s/n, 4169-007 Porto, Portugal}

	
	\author{Margarida Lima\orcidMargarida\!\!} \email{margarida.a.lima@tecnico.ulisboa.pt}
	\affiliation{OKEANOS–Instituto de Investigaç\~{a}o em Ci\^{e}ncias do Mar, Universidade dos Açores,
		Rua Prof. Doutor Frederico Machado, 4, 9900-140 Horta, Portugal}
	\affiliation{Departamento de F\'{i}sica, Instituto Superior T\'{e}cnico, Universidade de Lisboa,
		Av. Rovisco Pais, 1049-001 Lisboa, Portugal}
	\affiliation{Centro de An\'{a}lise Matem\'{a}tica, Geometria e Sistemas Din\^{a}micos, Departamento de Matemática, Instituto Superior T\'{e}cnico,
		Universidade de Lisboa, Av. Rovisco Pais, 1049-001 Lisboa, Portugal}
	
	\author{Francisco S. N. Lobo\orcidFrancisco\!\!} \email{fslobo@ciencias.ulisboa.pt}
	\affiliation{Instituto de Astrof\'{i}sica e Ci\^{e}ncias do Espa\c{c}o, Faculdade de Ci\^{e}ncias da Universidade de Lisboa, Edifício C8, Campo Grande, P-1749-016 Lisbon, Portugal}
	\affiliation{Departamento de F\'{i}sica, Faculdade de Ci\^{e}ncias da Universidade de Lisboa, Edif\'{i}cio C8, Campo Grande, P-1749-016 Lisbon, Portugal}
	
	
	\author{Luís F. Dias da Silva\orcidLuis\!\!} 
	\email{fc53497@alunos.fc.ul.pt}
	\affiliation{Instituto de Astrof\'{i}sica e Ci\^{e}ncias do Espa\c{c}o, Faculdade de Ci\^{e}ncias da Universidade de Lisboa, Edifício C8, Campo Grande, P-1749-016 Lisbon, Portugal}
	\affiliation{Departamento de F\'{i}sica, Faculdade de Ci\^{e}ncias da Universidade de Lisboa, Edif\'{i}cio C8, Campo Grande, P-1749-016 Lisbon, Portugal}

	\date{\LaTeX-ed \today}
	
	\begin{abstract}
		We study black hole shadows in nonminimally coupled Weyl connection gravity, a metric-affine extension of general relativity in which spacetime is described by a metric and a Weyl vector field encoding non-metricity. Despite going beyond the Riemannian framework, the presence of a non-dynamical Weyl vector ensures second-order field equations. The theory admits Schwarzschild- and Reissner--Nordstr\"{o}m-like solutions modified by a Weyl integration constant that parametrizes deviations from General Relativity.
		By computing the corresponding shadow radii and confronting them with the Event Horizon Telescope constraints on Sgr A*, we place observational bounds on the Weyl parameter. Assuming an observer distance $r_O = 4.1\times 10^{10}M$ and requiring consistency at the $2\sigma$ level, we obtain $\omega \gtrsim 10^{11.7}M$ (model I), $\omega \gtrsim 10^{10.5}M$ (model II), and $\omega \sim 10^{12}M$ (model III). 
		Our results show that present horizon-scale imaging already sets meaningful limits on spacetime non-metricity. This work highlights the power of black hole shadow observations as probes of extended gravitational dynamics and establishes a direct link between Weyl-based theories and current astrophysical data.
	\end{abstract}
	
	\maketitle
	
\section{Introduction}

We live in a multimessenger era in which observational data from a wide range of astrophysical and cosmological sources are available as never before. This wealth of information enables increasingly precise tests of the nature of gravity~\cite{CMWill}. At the same time, a growing number of alternative theories of gravity have been proposed that extend or modify Einstein's theory of General Relativity~\cite{Capozziello:2011,Odintsov:2011,Odintsov:2017}. It is therefore essential that these models be carefully scrutinized.
One of the simplest extensions of General Relativity is given by the so-called $f(R)$ theories, in which the Ricci scalar in the gravitational action is replaced by a generic function of $R$~\cite{fR1,fR2}. These models offer certain advantages, most notably the possibility of providing a natural explanation for the late-time accelerated expansion of the Universe, although they also face several limitations. Motivated by these issues, further generalizations have been proposed in which an additional function of the curvature is nonminimally coupled to the matter Lagrangian. Such models may account for galactic rotation curves without invoking exotic components such as dark matter~\cite{nmc}.

However, these theories typically assume a metric--\break compatible connection, namely the Levi--Civita connection, implying that the metric tensor alone fully characterizes the geometry of spacetime. In this case, the connection is uniquely determined by the metric and is both torsion-free and metric-compatible.
In a more general geometric framework, the affine connection can be treated as an independent object and may contain additional degrees of freedom. In particular, it may exhibit torsion~\cite{Cai:2015emx}, $T$, associated with the antisymmetric part of the connection, and non-metricity~\cite{BeltranJimenez:2017tkd,BeltranJimenez:2019tme,BeltranJimenez:2019esp}, $Q$, which quantifies the failure of the metric to be preserved under parallel transport. Accordingly, the affine connection can be decomposed into independent contributions associated with curvature, torsion, and non-metricity.
Remarkably, in four dimensions, gravitational actions constructed from the linear scalar invariant of each of these geometric quantities yield theories that are formally equivalent at the level of the field equations~\cite{BeltranJimenez:2019esp}. This correspondence highlights the equivalence between the curvature, teleparallel, and symmetric teleparallel formulations of General Relativity. However, such equivalence generally breaks down in extended theories, such as $f(R)$, $f(T)$, and $f(Q)$ models.

Weyl proposed a related geometric extension in 1918 in an attempt to unify electromagnetism with gravity~\cite{Weyl:1918}. In Weyl geometry, the metric compatibility condition is relaxed and the non-metricity is described by a vector field, which can naturally be interpreted as a gauge field associated with electromagnetism. In this picture, the electromagnetic interaction arises as a geometric property of spacetime.
This idea was later revisited by Dirac, who introduced an additional scalar field in order to preserve scale invariance within the theory~\cite{Dirac:1973}. Despite their conceptual appeal, these formulations suffer from well-known difficulties, most notably the so-called second clock effect, which leads to unphysical predictions for the behavior of clocks transported along different paths. Nevertheless, Weyl-inspired frameworks have attracted renewed interest in recent years, particularly in cosmology, where the additional geometric degrees of freedom may contribute to the description of dark components of the Universe or the dynamics of cosmic inflation~\cite{Alvarez:2016qfa}.

In this context, the nonminimal matter--curvature coupling framework has also been extended to the non-metricity sector, where the Ricci scalar is replaced by the scalar invariant constructed from the non-metricity part of the connection~\cite{Harko:2018gxr}. A further natural step is to incorporate Weyl's geometric idea into this class of models, leading to the so-called nonminimally coupled Weyl connection gravity~\cite{Gomes:2018weyl}. This theory admits a viable cosmological description~\cite{Baptista:2019}. Moreover, when the Weyl vector is identified with the electromagnetic gauge field and a Faraday-type matter Lagrangian is included, the theory can avoid Ostrogradsky instabilities provided that certain conditions are satisfied, such as a Weyl vector of the form $W_{\mu}=(0,W(r),0,0)$ or a vanishing extrinsic curvature scalar associated with the spacetime foliation~\cite{Baptista:2020}. 

The weak-field regime of the theory has also been analyzed and shown to be stable~\cite{Gomes:2025weakweyl}. In addition, black hole solutions and their thermodynamic properties have been obtained and characterized in this framework~\cite{Lima:2024cys}. The model further impacts scalar cosmological perturbations and leads to modifications of the Friedmann equations~\cite{Lima:2025}. 
More broadly, several related extensions involving the Weyl vector or the Weyl tensor have been proposed in recent years, including scenarios admitting black hole solutions~\cite{Kouniatalis:2024gnr,Yang:2022icz,Sakti:2024pze}.

In fact, black hole physics has also experienced remarkable progress in recent years. Advances range from the development of increasingly accurate black hole models used to generate templates for gravitational-wave signals, to large-scale numerical simulations such as the supercomputing effort that produced the first realistic visualization of a black hole for the film \textit{Interstellar}~\cite{James:2015yla}. Most notably, the Event Horizon Telescope has provided the first direct images of black hole shadows, including the supermassive black hole at the center of the galaxy M87~\cite{EventHorizonTelescope:2019ths} and the black hole at the center of our Galaxy, Sgr~A*~\cite{EventHorizonTelescope:2022wok}. These observations offer an unprecedented opportunity to probe the strong-field regime of gravity.

Motivated by these developments, in this work we investigate black hole shadows in the framework of nonminimally coupled Weyl connection gravity. In particular, we analyze the properties of the shadow and compare the resulting normalized shadow radius with observational constraints from Sgr~A*. This analysis provides insight into the interplay between the metric and the Weyl vector field, and allows us to explore possible deviations from General Relativity, including the potential presence of a cosmological horizon in addition to the event horizon.

This paper is organized as follows. In Sec.~\ref{sec:Model}, we introduce the nonminimally coupled Weyl gravity model. In Sec.~\ref{sec:BHModels}, we review the Schwarzschild- and Reissner--Nordström-like black hole solutions arising in this framework. In Sec.~\ref{sec:BHshadows}, we analyze the corresponding black hole shadows and compare the predicted normalized shadow radius with the observational constraints from Sgr~A*. Finally, our conclusions are presented in Sec.~\ref{sec:conc}.

\section{Nonminimally coupled Weyl connection gravity model}\label{sec:Model}

The nonminimally coupled Weyl connection gravity model is described by the action functional \cite{Gomes:2018weyl}:
\begin{equation}
    S=\int \left( \kappa f_1(\bar{R})+f_2(\bar{R}) \mathcal{L} \right)\sqrt{-g}~\mathrm{d}^4\mathrm{x},\label{NMC-action}
\end{equation}
where $g$ denotes the determinant of the metric tensor $g_{\mu\nu}$, $\mathcal{L}$ is the Lagrangian density of the matter fields, and $\kappa=(16\pi G)^{-1}$, with $G$ being Newton's gravitational constant. In what follows, and without loss of generality, we set $\kappa=1$. The functions $f_1(\bar{R})$ and $f_2(\bar{R})$ are generic functions of the generalized curvature scalar $\bar{R}=R+\bar{\bar{R}}$, which contains two contributions: the Ricci scalar $R$ and an additional scalar $\bar{\bar{R}}$ associated with non-metricity.

In this framework, the affine connection is decomposed into two parts: the Levi--Civita connection and a contribution associated with non-metricity (for a systematic classification of general affine connections see, e.g., Ref.~\cite{Castillo-Felisola:2025zrh}). The latter is given by
\begin{equation}
	\bar{\bar{\Gamma}}^{\rho}{}_{\mu\nu}
	=
	-\frac{1}{2}\delta^{\rho}_{\mu}W_{\nu}
	-\frac{1}{2}\delta^{\rho}_{\nu}W_{\mu}
	+\frac{1}{2}g_{\mu\nu}W^{\rho},
\end{equation}
where $W_{\mu}$ is the Weyl vector defined through the non-metricity condition $D_{\lambda} g_{\mu\nu} = W_{\lambda} g_{\mu\nu}$, with $D_{\lambda}$ denoting the generalized covariant derivative constructed from the full connection. Throughout this work, $\nabla_{\lambda}$ will denote the covariant derivative associated with the Levi--Civita connection.

In the present formulation, the affine connection is not an independent field in the Palatini sense; it is assumed to be the Weyl connection determined by the metric and the Weyl vector. Consequently, the variational principle is applied only to the metric and the Weyl vector, not to the connection separately. This is the most general variation consistent with the Weyl geometric structure.
	However, one might ask what would change if the connection were treated as an
	independent field in the usual Palatini sense, i.e., varying the action with respect to $g_{\mu\nu}$, $W_\mu$, and the full connection
		$\Gamma^\rho_{\mu\nu}$ independently. For a general action of the form
		$S = \int \big(\kappa f_1(\bar R) + f_2(\bar R)\mathcal{L}\big)\sqrt{-g}$,
		a Palatini variation would impose additional constraints from the
		connection equations. In the absence of torsion and with the non-metricity
		tensor defined as $Q_{\lambda\mu\nu} = D_\lambda g_{\mu\nu}$, the
		connection equations typically force the connection to be the
		Levi-Civita connection of an auxiliary metric, or to satisfy metric
		compatibility and zero torsion. For the specific case of the Weyl
		connection ansatz (where non-metricity is proportional to $W_\lambda g_{\mu\nu}$), a full Palatini variation would impose additional constraints that are generally incompatible with a non‑zero Weyl vector. One would
		find that the Weyl vector must vanish or that the theory collapses to
		standard general relativity (or a scalar-tensor theory) because the
		connection equations force the non-metricity to be zero. Hence, the
		present formulation, where the connection is fixed to be the Weyl
		connection from the outset, is necessary to maintain a non-trivial
		non-metricity degree of freedom. This approach avoids the
		over-constraints of the Palatini treatment and allows the Weyl vector
		to play an active geometric role.

From this generalized connection, one can construct the corresponding extensions of the Riemann tensor and its contractions. In particular, the generalized Ricci tensor is given by:
\begin{eqnarray}
\bar{R}_{\mu\nu} & = & R_{\mu\nu}+\frac{1}{2}W_\mu W_\nu +\frac{1}{2}g_{\mu\nu} \left(\nabla_\lambda-W_\lambda\right)W^\lambda +F_{\mu\nu}
	\nonumber\\
&&+\frac{1}{2}\left(\nabla_\mu W_\nu+\nabla_\nu W_\mu\right):=R_{\mu\nu}+\bar{\bar{R}}_{\mu\nu},
	\label{Riccitensor}
\end{eqnarray}
where $F_{\mu\nu}=\nabla_\mu W_\nu-\nabla_\nu W_\mu$ is the field-strength tensor associated with the Weyl vector. Its contraction then leads to a generalized curvature scalar:
\begin{equation}
	\bar{R}=R+3\nabla_\lambda W^\lambda-\frac{3}{2}W_\lambda W^\lambda:=R+\bar{\bar{R}}.
    \label{RicciScalar}
\end{equation}

Varying the action with respect to the Weyl vector, we obtain:
\begin{equation}
 	\nabla_\lambda \Theta(\bar{R})=-W_\lambda \Theta(\bar{R})~,
 	\label{eqconstraint}
 \end{equation}
 which is a constraint-like equation, and where $\Theta(\bar{R}):=F_1(\bar{R})+F_2(\bar{R})\mathcal{L}$, with $F_i(\bar{R}):=df_i(\bar{R})/d\bar{R}$, with $i~\in~\{1,2\}$.
 
Varying the action with respect to the metric tensor and using the previous result, we obtain:
 \begin{eqnarray}
    \bar{R}_{(\mu\nu)}\Theta(\bar{R})-\frac{1}{2}g_{\mu\nu}f_1(\bar{R})=\frac{1}{2}f_2(\bar{R})T_{\mu\nu}, \label{NMC-field-eqs}
\end{eqnarray}
where $\bar{R}_{(\mu\nu)}=R_{\mu\nu}+\bar{\bar{R}}_{(\mu\nu)}$ and $T_{\mu\nu}$ is the standard energy-momentum tensor constructed from the matter Lagrangian density.

Taking the divergence of the metric field equations and employing the Bianchi identities, one obtains a generalized non-conservation law for the energy-momentum tensor of the matter fields:
\begin{align}
	\nabla_\mu T^{\mu\nu} 
	&= \frac{2}{f_2(\bar{R})} \Bigg[
	\frac{F_2(\bar{R})}{2} \big( g^{\mu\nu} \mathcal{L} - T^{\mu\nu} \big) \nabla_\mu R
	 \nonumber \\
	& \hspace{-1.4cm} + \nabla_\mu \big( \Theta(\bar{R}) B^{\mu\nu} \big) - \frac{1}{2} \big( F_1(\bar{R}) g^{\mu\nu} + F_2(\bar{R}) T^{\mu\nu} \big) 
	\nabla_\mu \bar{\bar{R}} 
	\Bigg],
	\label{nonconserveq}
\end{align}
where the tensor $B^{\mu\nu}$ is defined by
\begin{equation}
	B^{\mu\nu} = \frac{3}{2} W^\mu W^\nu + \frac{3}{2} g^{\mu\nu} (\nabla_\lambda - W_\lambda) W^\lambda.
\end{equation}

This gravitational model has notable implications for astrophysics and cosmology. Unlike the metric formulation of the nonminimal matter--curvature coupling model, it does not introduce higher-order derivatives in the metric field equations and features a non-dynamical Weyl vector field. Furthermore, the energy-momentum tensor is generally non-conserved, as it depends explicitly on both the choice of matter Lagrangian and the configuration of the Weyl vector.

\section{Black hole models}\label{sec:BHModels}

Black hole solutions in this model were derived in Ref.~\cite{Lima:2024cys}, including those analogous to the Schwarzschild and \break Reissner--Nordström solutions of General Relativity. Here, we summarize the relevant solutions to facilitate the analysis of the corresponding black hole shadows.

To ensure that the lengths remain positive under parallel transport, it was required that $W_{\lambda} dx^{\lambda} \geq 0$. Moreover, the spacetime metric can be written in the general form
\begin{equation}
ds^2=-g_{tt}dt^2+g_{rr}dr^2+r^2(d\theta^2+\sin^2(\theta)d\phi^2).
\end{equation}

Due to the static nature of the problem and the symmetries of the spacetime metric, two distinct forms of the Weyl vector can be considered, namely: $W_\mu=(0,W(r),0,0)$, with $W(r)>0$ and  $W_\mu=\left(W_0(r),W_1(r),0,0\right)$, such that $W_0'(r) + \left(W_1(r) - \frac{g_{tt}'}{g_{tt}}\right) W_0(r) = 0$.

The condition relating $W_0$ and $W_1$ is obtained from the requirement that the generalized Ricci tensor $\bar{R}_{(\mu\nu)}$ vanish, which is necessary for vacuum solutions. The explicit derivation is given in Ref.~\cite{Lima:2024cys}. Physically, this condition ensures that non-metricity does not introduce unwanted anisotropies, preserving spherical symmetry.

These Weyl vector configurations are used to construct the corresponding black hole metrics and analyze the resulting shadows. The first ansatz is particularly convenient for spherically symmetric solutions, while the second allows for more general configurations consistent with the field equations.

\subsection{Schwarzschild-like black hole}

We now focus on Schwarzschild-like black hole solutions, corresponding to non-rotating and uncharged configurations characterized by the local mass parameter $M$. 

The vacuum field equations impose the conditions \break $\bar{R}_{\mu\nu}=0$ and $\bar{R}=0$, where these generalized quantities include contributions from both the metric and the Weyl vector. Under these conditions, the Weyl vector may assume the two distinct forms introduced above.

\subsubsection{First case: $W_{\mu}=(0, W(r), 0, 0)$}\label{First-Case}

For the first configuration, the field equations imply that
\begin{equation}
    W(r) = \frac{2}{r + \omega},
\end{equation}
where $\omega > 0$ is a constant, here referred to as Weyl's constant. The corresponding metric components are then given by
\begin{equation}\label{eq:model_I}
    g_{tt}=1-\frac{2 M}{r} + \frac{2(\omega + 3 M)}{\omega}\frac{r}{\omega} + \frac{\omega + 4 M}{\omega} \left( \frac{r}{\omega} \right)^2.
\end{equation}
and
\begin{equation}
g_{rr}=\frac{1+6(M/\omega)}{g_{tt}}~.
\end{equation}

In this black hole solution, a single event horizon appears, located at $r_H=2M \omega/(4M+\omega)$. This is a vacuum solution that, despite $\bar{R}=0$, has a nonzero value for the Ricci scalar, given by
\begin{equation}
	R=-\frac{12(r+\omega)\left( (4M+\omega)r-M\omega\right)}{\omega^2(6M+\omega)r^2}.
	\label{Ricci-sol-CASE1}
\end{equation}

	The coefficient of the $r^2$ term in Eq.~\eqref{eq:model_I} can be interpreted as an effective cosmological constant $\Lambda_{\text{eff}} = 3(\omega+4M)/\omega^3$. Observational constraints from cosmology (e.g., the measured dark energy density) would imply $\omega \sim 10^{61}$ in Planck units, which is many orders of magnitude larger than the shadow bounds. This indicates that the quadratic term cannot be identified with the physical cosmological constant unless the Weyl constant is fine‑tuned.

\subsubsection{Second case: $W_{\mu}=(W_0(r),W_1(r),0,0)$}\label{Second-Case}

In this second case, the field equations require that
\begin{equation}
    W_0(r)=\frac{1}{\omega}\left( 1-\frac{2M}{r}  \right) \quad\text{and}\quad W_1(r)=\frac{2r}{r^2-4\omega^2},
\end{equation}
with $\omega>0$. For this, we get the following solution for the metric component:
\begin{equation}\label{eq:model_II}
    g_{tt}=1-\frac{2 M}{r} + \frac{M}{2\omega}\left(\frac{r}{\omega}\right) - \frac{1}{4} \left( \frac{r}{\omega} \right)^2,
\end{equation}
and $g_{rr}=1/g_{tt}$. 

This solution admits two possible event horizons, located at 
$r^{^{(M)}}_{H}~=~2M$ and $r^{^{(\omega)}}_{H}=2 \omega$.
Analogously to the previous case, although the generalized curvature scalar vanishes, the Ricci scalar is nonzero and is given by
\begin{equation}
	R=\frac{3(r-M)}{\omega^2 r}.\\
\end{equation}

	For Model II, the $r^2$ term gives an effective cosmological constant $\Lambda_{\text{eff}} = -3/(4\omega^2)$, which would be negative (de Sitter‑like). Cosmological constraints again demand $\omega$ to be extremely large, making the shadow corrections negligible.

The two solutions discussed in Secs.~\ref{First-Case} and \ref{Second-Case} are derived in vacuum. However, one can also consider a scenario where the matter content is represented by a cosmological constant, with the Lagrangian density
$\mathcal{L}^{(\Lambda)} = -2\Lambda$.
It has been shown that the vacuum solutions remain valid even in the presence of a cosmological constant. This can be interpreted either as a straightforward mathematical reparametrization or as the inclusion of a physical vacuum energy term. While these two perspectives are mathematically equivalent, they lead to distinct physical interpretations: in the first case, the cosmological constant appears merely as a coefficient in the reparametrization, whereas in the latter, it is associated with the vacuum energy of spacetime.

\subsection{Reissner--Nordstr\"{o}m-like black hole}

In this section, we review a solution describing Reissner--Nordstr\"{o}m black holes, i.e., static black holes characterized by mass $M$ and electric charge $Q$.

To model the gravitational field outside a charged, non-rotating, spherically symmetric body, the electrostatic four-potential is assumed to take the form \break $\Phi_\mu = (-\phi(r), 0, 0, 0)$, where $\phi(r)$ is the scalar potential. Consistency with the symmetry of the system further restricts the Weyl vector to a purely radial form.

It was shown that the Reissner--Nordstr\"{o}m solution does not exist in vacuum within this framework. Therefore, to obtain the first non-trivial modifications to the geometry, the matter content was modeled as a cosmological constant.

The field equations require $W(r)=2/(r+\omega)$ and $\phi(r)=\tilde{Q}/r$, where $\omega>0$ and $\tilde{Q}$ represent the dressed charge, which has units of $M$ and is related to the usual charge $Q$ via
\begin{equation}\label{eq:dressed_charge}
	Q^2=\zeta^2 \tilde{Q}^2 \left( 1+6 \left(\frac{M}{\omega}+\frac{\tilde{Q}^2}{\omega^2} \right) \right)^{-1},
\end{equation}
with $\zeta$ a constant. 

The scalar potential $\phi(r)=\tilde{Q}/r$ exhibits the standard Coulomb decay $1/r$, but with an effective strength modified by the parameter $\zeta$ through the relation between $Q$ and $\tilde{Q}$. In the asymptotically flat limit ($\omega\to\infty$), $\phi(r) = Q/(\zeta r)$, so the potential reduces to the usual Reissner–Nordström form up to a rescaling.

For this, we get the following solution for the metric components: 
\begin{equation}
	\begin{aligned}\label{eq:model_III}
		g_{tt}= 1
		& - \frac{2 M}{r} +\frac{\tilde{Q}^2}{r^2} +\frac{2(\omega^2 + 3M\omega + 2\tilde{Q}^2)}{\omega^2}\frac{r}{\omega} \\
		& +\frac{\omega + 4 M}{\omega} \left( \frac{r}{\omega} \right)^2,
	\end{aligned}
\end{equation}
and
\begin{equation}
	g_{rr}=\frac{1+6\left(\frac{M}{\omega}+\frac{\tilde{Q}^2}{\omega^2}\right)}{g_{tt}}. 
\end{equation}

	The physical origin of the charge in these solutions can be traced to the matter Lagrangian. Including a Maxwell term $-\frac14 F_{\mu\nu}F^{\mu\nu}$ in $\mathcal{L}$ yields a Coulomb‑type solution. The nonminimal coupling $f_2(\bar R)\mathcal{L}$ renormalizes the charge, giving rise to the dressed charge $\tilde Q$ and the correction factor $\zeta$. Charge can therefore be induced by accretion of charged matter or quantum pair production in the strong gravitational field.

Depending on the values chosen for the parameters, this solution may exhibit two, one, or no event horizons.  

In this scenario, the generalized curvature scalar takes the form
\begin{equation}
	\bar{R}=\frac{36 \tilde{Q}^2}{\left( \omega^2+6M\omega+6\tilde{Q}^2 \right)\left(r+\omega\right)^2}.
	\label{RN-curvature-scalar}
\end{equation}

Likewise, the Ricci scalar is given by
\begin{equation}
	\begin{split}
		R = & -\frac{12}{r^2 \, \omega \left( \omega^2 + 6 M \omega + 2 \tilde{Q}^2 \right)} 
		\Big[  (4M+\omega) r^2 \\
		 & + (\omega^2 + 3 M \omega + 2 \tilde{Q}^2) r  - (\tilde{Q}^2 + M \omega) \, \omega \Big] .
	\end{split}
	\label{RN-Ricci}
\end{equation}

\section{Black hole shadows}\label{sec:BHshadows}

The term \textit{shadow} refers to the region on an external observer's sky bounded by the critical curve, which separates null geodesics captured by the black hole from those that escape to infinity~\cite{Gralla:2019xty}. This curve corresponds to the gravitationally lensed projection of the black hole's unstable photon region~\cite{Falcke:1999pj,Bronzwaer:2021lzo,Perlick:2021aok}. Because the shadow's properties are sensitive to the spacetime geometry~\cite{Cunha:2018acu,Chen:2022scf}, it can be used to perform null-tests of black hole models against current observations~\cite{Cardoso:2019rvt}.

In the following, we briefly summarize the derivation of the shadow radius for a generic static, spherically symmetric, and non-asymptotically flat spacetime (see Ref.~\cite{Perlick:2021aok} for a detailed review of analytical calculations of black hole shadows). We consider a general static and spherically symmetric metric of the form
\begin{equation}
    ds^2=A(r) dt^2 - B(r) dr^2 - C(r) d\Omega^2,\label{eq:generic_metric}
\end{equation}
where $A(r), B(r)$ and $C(r)$ denote generic radially dependent functions, and $d\Omega^2 = d\theta^2 + \sin{\theta}d\phi^2$. We can leverage the spherical symmetry of the problem and restrict the analysis to equatorial plane geodesics $\theta=\pi/2$, without loss of generality. 
In such cases, the equation of motion along null geodesics may be expressed as
\begin{equation}
    \left(\frac{dr}{d\phi} \right)^2=\frac{C(r)^2}{A(r)B(r)}\left( \frac{1}{b^2}-\frac{A(r)}{C(r)}\right),\label{eq:photon_motion}
\end{equation}
where $b$ denotes the geodesic trajectory impact parameter. 

Consider a static observer located at some distance $r_O$ from a black hole with mass $M$, and a light ray, traced backwards from the observer to its origin, at an angle $\alpha$ relative to the radial coordinate $r$. One can relate the angle $\alpha$ with the geometry via
\begin{equation}
    \cot\alpha=\sqrt{\frac{B(r)}{C(r)}}\frac{dr}{d\phi}\Bigg|_{r=r_{O}}.\label{eq:lightray_angle}
\end{equation}
Substituting Eq.~\eqref{eq:photon_motion} into Eq.~\eqref{eq:lightray_angle} yields, after some \break trigonometric manipulation,  the following relation between the angular size and the impact parameter
\begin{equation}
    \sin^2\alpha=b^2 \frac{A(r)}{C(r)}\Bigg|_{r=r_{O}}.\label{eq:lightray_angle_2}
\end{equation}

The perimeter of the shadow is associated with the critical curve, defined by the critical value of the impact parameter for photon trajectories that asymptotically approach the family of unstable bound geodesics \cite{Gralla:2019xty}. These trajectories satisfy $dr/d\phi=0$, and therefore Eq.~\eqref{eq:photon_motion} yields	
\begin{equation}\label{eq:impact_param}
    b^2_{cr}=\frac{C(r_{ph})}{A(r_{ph})},
\end{equation}
where $r_{ph}$ is the radius of the unstable photon region, that is a solution of \cite{Claudel:2000yi}
\begin{equation}
    \frac{C^{\prime}(r_{ph})}{C(r_{ph})} = \frac{A^{\prime}(r_{ph})}{A(r_{ph})}\,.
    \label{eq:photon_sphere}
\end{equation}

Inserting Eq. \eqref{eq:impact_param} into Eq. \eqref{eq:lightray_angle_2}, in the small angle approximation (i.e., valid at $r_O \gg M$), returns the formula for the angular shadow radius
\begin{equation}
    \alpha^2_{sh}=\frac{C(r_{ph})}{A(r_{ph})} \frac{A(r)}{C(r)}\Bigg|_{r=r_{O}},\label{eq:angular_shadow_radius}
\end{equation}
which, in the small angle approximation (i.e., valid at $r_O \gg M$), provides the shadow radius
\begin{equation}
r_{sh}=r_{ph}\sqrt{\frac{A(r_{O})}{A(r_{ph})}},\label{eq:shadow_radius}
\end{equation} 
for all non-asymptotically flat spacetimes where $C(r)=r^2$ .


To analyze the proposed geometries, we adopt a strategy based on confronting the theoretically predicted shadow radius of each spacetime with the observational constraints on the shadow of Sgr A* reported by the EHT collaboration. Because the instrument lacks sensitivity to radiation below a certain fraction of the peak brightness, a direct measurement of the shadow radius of Sgr A* is not feasible. However, the EHT team argues that this obstacle can be overcome by employing the angular diameter of the surrounding luminous emission ring as a surrogate for the shadow size, once the relevant systematic biases are properly taken into account (see the detailed discussion in \cite{EventHorizonTelescope:2022urf}). Owing to previous independent determinations of the mass-to-distance ratio $M/D$\footnote{The parameters $M$ and $D$ for Sgr A* were inferred from high-precision monitoring of the orbits of the S-stars, which belong to the nuclear stellar cluster at the Galactic center (for further details, see \cite{EventHorizonTelescope:2022urf}).} by the Keck Observatory \cite{Do:2019txf} and the Very Large Telescope Interferometer (VLTI) \cite{GRAVITY:2020gka} collaborations, the EHT analysis constrains the fractional difference $\delta$ between the reconstructed shadow diameter and that expected for a Schwarzschild black hole, whose angular scale is $\theta_{sh,Sch}=6\sqrt{3}\, M/D$, expressed as \cite{EventHorizonTelescope:2022urf}
\begin{equation}\label{eq:fractional_deviation}
    \delta = \frac{r_{sh}}{3\sqrt{3}M} - 1 \ .
\end{equation}

Accordingly, this fractional discrepancy may be translated into quantitative constraints on the measured shadow radius of Sgr A*, evaluated at the $1\sigma$ confidence level
\begin{equation} \label{eq:1sigma}
    4.55 \lesssim r_{s}/M \lesssim 5.22 \ ,
\end{equation}
and at $2\sigma$
\begin{equation} \label{eq:2sigma}
    4.21 \lesssim r_{s}/M \lesssim 5.56 \ .
\end{equation}
These constraints are derived by taking the mean of the fractional deviations obtained using the Keck and VLTI determinations of $M/D$, while modeling the associated errors with a Gaussian distribution \cite{Vagnozzi:2022moj}. In what follows, we also adopt the mean of the distance estimates reported by the Keck and VLTI collaborations for the observer’s location, which in geometrized units corresponds to approximately $r_O = 4.1\times 10^{10}M$.

We emphasize that this procedure is subject to two primary limitations. First, in order to assess the theoretical systematics affecting the calibration factor employed in extracting the shadow size of Sgr A*, the EHT collaboration relies on synthetic black hole images designed to probe a broad spectrum of accretion flow configurations and background geometries. In doing so, the Kerr metric is adopted as the reference framework, although the analysis also explores non-Kerr black holes and other compact-object scenarios, including wormholes and boson stars. Since the methodology is fundamentally grounded on asymptotically flat solutions within general relativity, geometries such as the Cotton spacetime—and, more generally, models that depart significantly from the GR paradigm—are not incorporated. Consequently, by applying the reported constraints to the CG spacetime, we inevitably introduce a degree of theoretical bias, as we presume the robustness of those results in this extended setting. A dedicated assessment of how Weyl-type geometries might alter the inferred fractional deviation, and hence the associated bounds, would require a dedicated study along the lines of the EHT analysis, which is beyond the scope of the present work.

The second caveat pertains the non-asymptotically flat nature of these spacetimes considered throughout this work, for which the Arnowitt-Deser-Misner (ADM) mass \cite{Arnowitt:1962hi} is ill-defined for an observer located at spatial infinity. Consequently, it cannot be directly associated with the mass parameter in the metric. In order to avoid this ambiguity, we adopt, as in \cite{Vagnozzi:2022moj}, a normalization of the shadow radius by the local mass parameter $M$, the same quantity constrained by the precise orbital tracking of the S-stars under the local gravitational field of Sgr A*.

\begin{figure}[ht!]
    \includegraphics[width=\columnwidth]{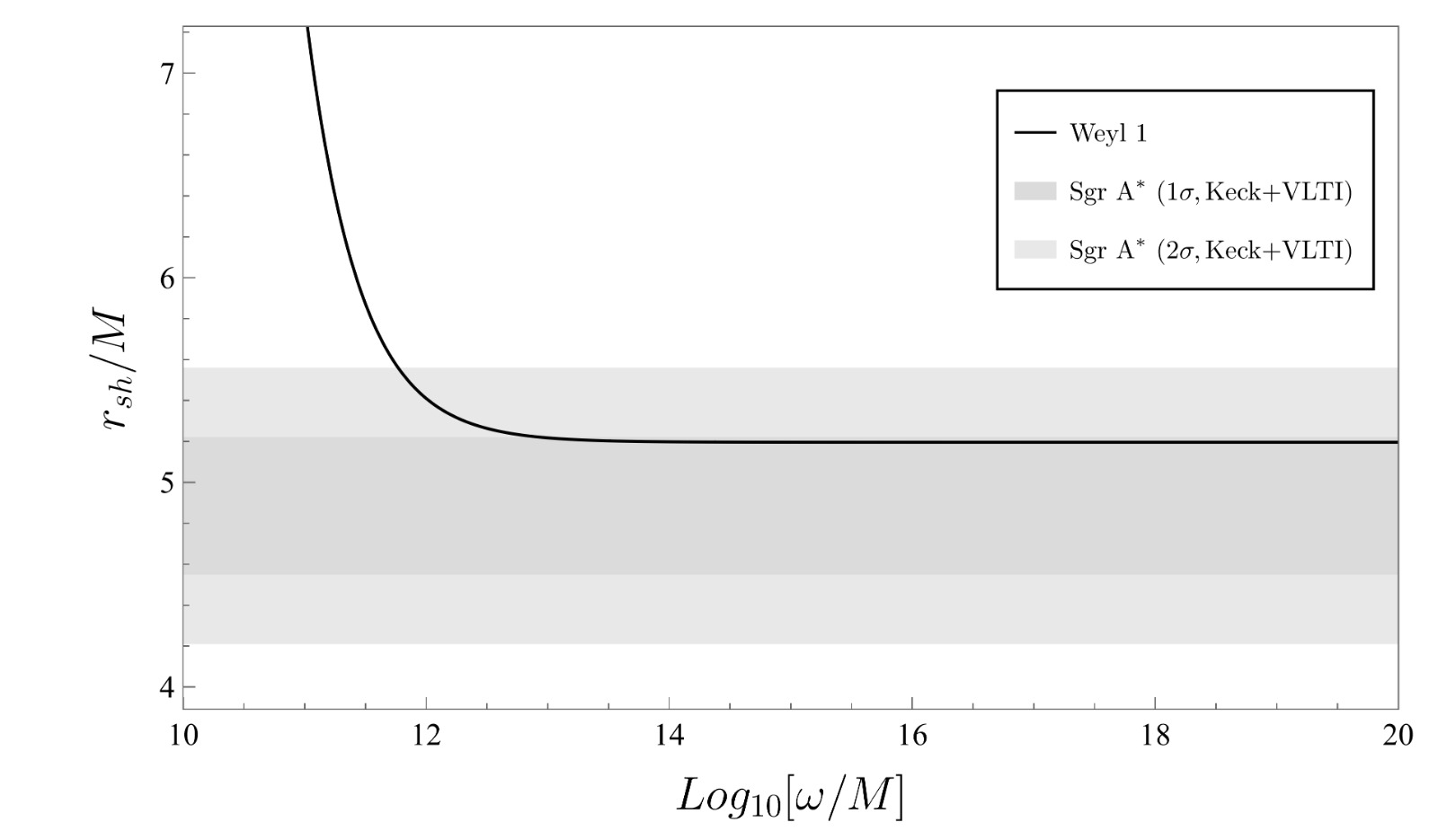}
    \caption{Shadow radius $r_{sh}$ of model I \eqref{eq:model_I} as a function of the Weyl constant $\omega$ (black line), for an observer at a distance of $r_O =  4.1\times 10^{10}M$. The shaded areas represent the confidence intervals of Sgr A*'s shadow radius at 1$\sigma$ (dark gray) and at 2$\sigma$ (light gray). The horizontal line $3\sqrt{3}M$ ($\sim 5.2$) marks the Schwarzschild shadow radius (see \cite{Perlick:2021aok}) for reference.}
    \label{fig:shadow_I}
\end{figure}

In Fig. \ref{fig:shadow_I}, we illustrate the dependence of the shadow radius on the Weyl constant $\omega$ in units of mass $M$, considering an observer located at $r_O = 4.1\times 10^{10}M$, for the spacetime model described by \eqref{eq:model_I}.

Observing Fig. \ref{fig:shadow_I} reveals that the theoretical shadow radius behaves inversely proportional to $\omega$: $r_{sh}$ decreases from infinity before converging to $3\sqrt{3}M$, where it coincides with the Schwarzschild black hole's shadow radius. This result is unsurprising, since $g_{tt} \approx 1 - 2 M/r$ at $\omega \rightarrow \infty$, yielding no detectable imprints in the shadow size. The turning point occurs at $\omega \sim 10^{12}$ because, at the considered observer's distance, the product $\omega r_O$ becomes non-negligible causing $r_{sh}$ to grow rapidly for lower values of $\omega$. Nonetheless, the EHT observational data constrains the parameter space of the Weyl constant in Model I to $\omega \gtrsim 10^{11.7} M$, where the theoretical shadow radius of model I remains compatible with the $2\sigma$ interval. 

\begin{figure}[ht!]
    \includegraphics[width=\columnwidth]{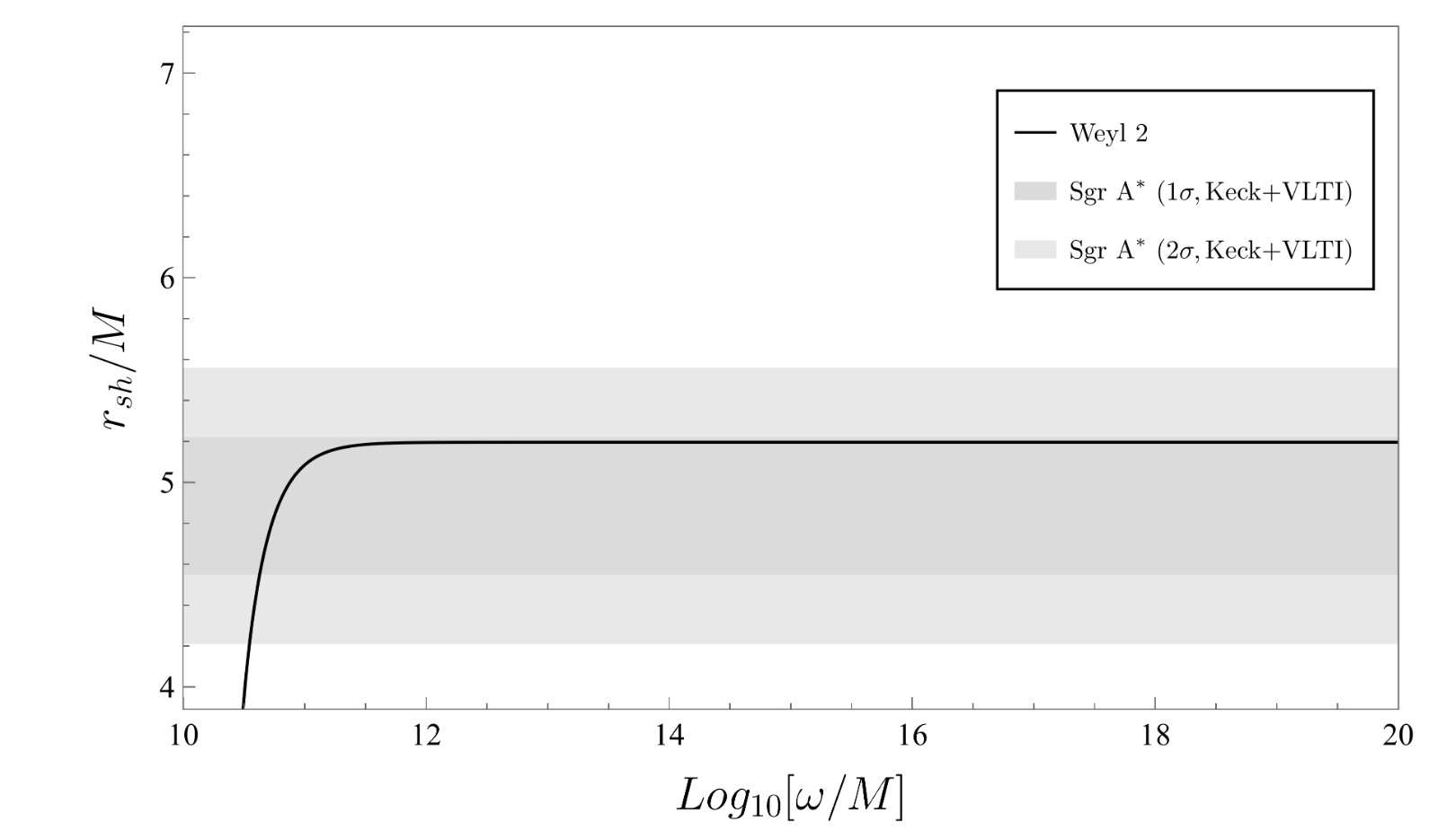}
    \caption{Shadow radius $r_{sh}$ of model II \eqref{eq:model_II} as a function of the Weyl constant $\omega$ (black line), for an external observer at $r_O =  4.1\times 10^{10}M$. The shaded areas represent the confidence intervals of Sgr A*'s shadow radius at 1$\sigma$ (dark gray) and at 2$\sigma$ (light gray). The horizontal line indicates the Schwarzschild shadow radius $3\sqrt{3}M$ (see \cite{Perlick:2021aok}) for comparison.}
    \label{fig:shadow_II}
\end{figure}

Similarly, Fig. \ref{fig:shadow_II}, shows the shadow radius as a function of $\omega$, as seen by an observer embedded in the geometry described by Eq. \eqref{eq:model_II} at $r_O = 4.1\times 10^{10}M$.

In this case, Fig. \ref{fig:shadow_II} shows that the shadow radius increases with $\omega$ up to a turning point near $\omega \sim  10^{11} M$, beyond which it asymptotes to the Schwarzschild shadow radius. Due to this behavior and the lower $2\sigma$ bound lying further from the canonical $3\sqrt{3}M$ radius, the constraints on $\omega$ are less stringent than in the previous case, namely $\omega \gtrsim 10^{10.5} M$.

\begin{figure}[ht!]
    \includegraphics[width=\columnwidth]{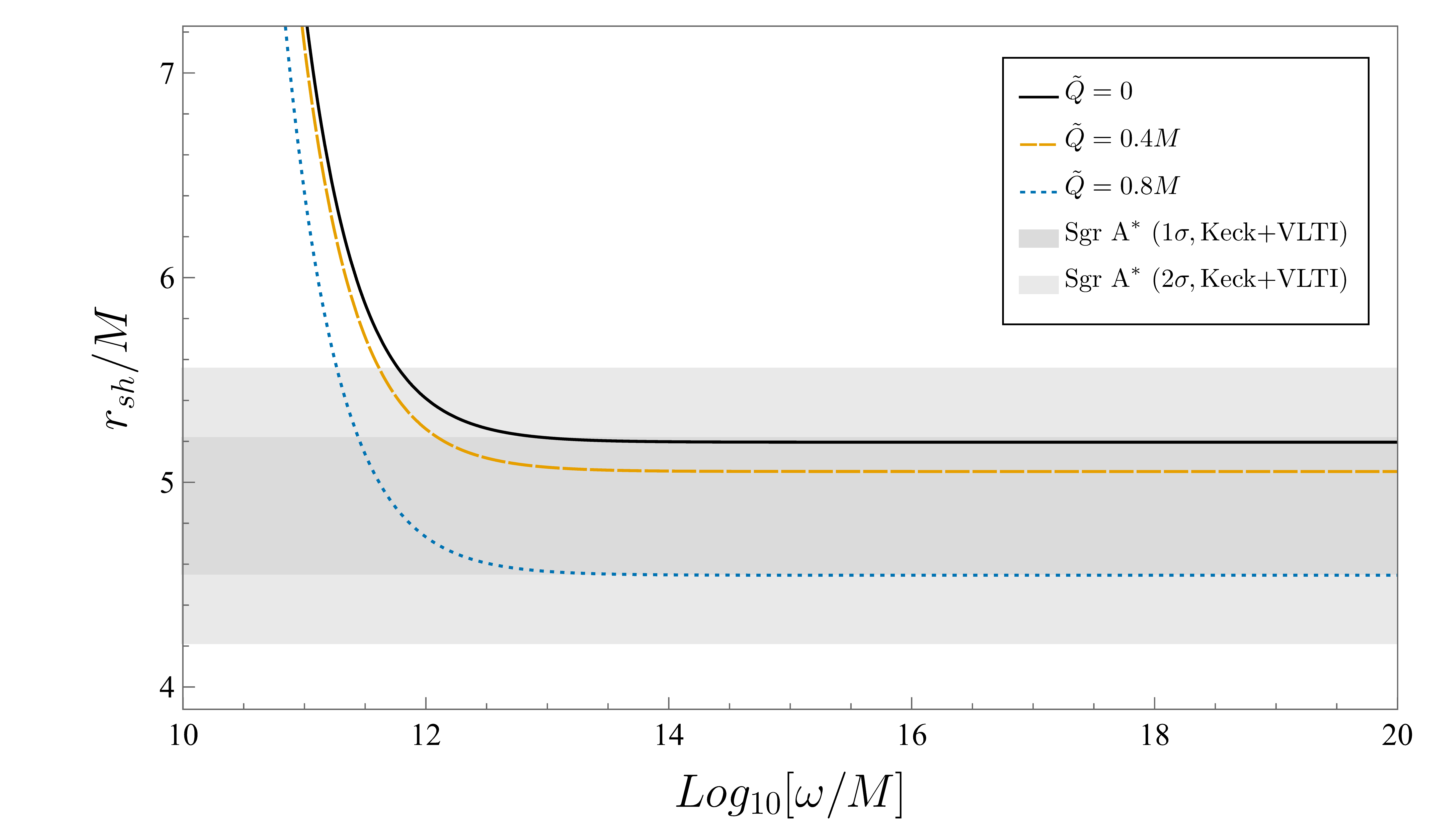}
    \caption{Shadow radius $r_{sh}$ of model III \eqref{eq:model_III} as a function of the Weyl constant $\omega$, for three dressed charge values: $\tilde{Q}=0$ (black solid line), $\tilde{Q}=0.4M$ (gold dashed line), $\tilde{Q}=0.8 M$ (dotted blue line); considering an external observer at $r_O =  4.1\times 10^{10}M$. The shaded areas represent the confidence intervals of Sgr A*'s shadow radius at 1$\sigma$ (dark gray) and at 2$\sigma$ (light gray). The horizontal  line shows the Schwarzschild shadow radius $3\sqrt{3}M$ (see \cite{Perlick:2021aok}).}
    \label{fig:shadow_III}
\end{figure}

Lastly, Fig. \ref{fig:shadow_III} shows the shadow radius of the space-time geometry \eqref{eq:model_III} as a function of $\omega$, considering an observer located at $r_O = 4.1\times 10^{10}M$. Because this particular model contains an additional free parameter, the dressed charge $\tilde{Q}$, we simultaneously study three scenarios of the latter.

As can be seen in Fig.~\ref{fig:shadow_III}, the relation between the shadow radius and the Weyl constant is qualitatively similar to that of the first model. Notably, the value of $r_{sh}$ is inversely proportional to that of $\omega$. Near $\omega \sim 10^{12} M$, a turning point occurs and the shadow converges to a specific value. For $\tilde{Q}=0$, it coincides with the Schwarzschild radius. For $\tilde{Q}=0.4M$ and $\tilde{Q}=0.8M$, the shadow matches that of a Reissner--Nordstr\"{o}m (RN) black hole with usual charge $Q=0.4M$ and $Q=0.8M$.  These results occur due to $g_{tt} \approx 1 - \frac{2 M}{r} + \frac{\tilde{Q^2}}{r^2}$ (i.e. RN black hole) in the limit $\omega \rightarrow \infty$. With regard to the charge, the shadow tends to grow smaller as the charge increases; a result which matches others already obtained in literature (compare with the first case studied in \cite{Vagnozzi:2022moj}). Given this behavior, the observational constraints to the parameter space of $\omega$ start at $\omega = 10^{11.7}M$, when $\Tilde{Q}=0$, and become more relaxed as the dressed charge increases, up until the value where the corresponding RN black hole shadow saturates the lower $2\sigma$ bound. Beyond that limit, the value of $\omega$ will remain confined to a smaller interval, as long as the geometry still casts a shadow. 

It is worth noting that the previous analysis takes into consideration the dressed charge $\tilde{Q}$, as opposed to the usual charge $Q$. Encoded in this definition lies an additional parameter, $\zeta$, which emerges as an integration constant and should have a detectable observational imprint. To study this parameter, we focus on the asymptotically flat regime ($\omega \rightarrow \infty$), for which Eq.~(\ref{eq:dressed_charge}) simplifies to $Q^2=\zeta^2 \tilde{Q^2}$ and the metric component to
\begin{equation}\label{eq:model_III_flat}
	g_{tt}= 1 - \frac{2 M}{r} +\frac{Q^2}{\zeta^2 r^2}
\end{equation}

Mathematically, the previous equation is analogous to a Reissner--Nordstr\"{o}m (RN) black hole geometry with a correction factor introduced in the charge. Qualitatively, for any value of the usual charge $Q$, one may find a value of $\zeta$ and $\tilde{Q}$ to reproduce the same results. However, the aim of the following analysis is to demonstrate that the emergence of this correction factor, for which there are currently no well-motivated theoretical constraints, either suppresses or enhances the size of the shadow and the optical appearance of the compact geometry.

\begin{figure}[ht!]
    \includegraphics[width=\columnwidth]{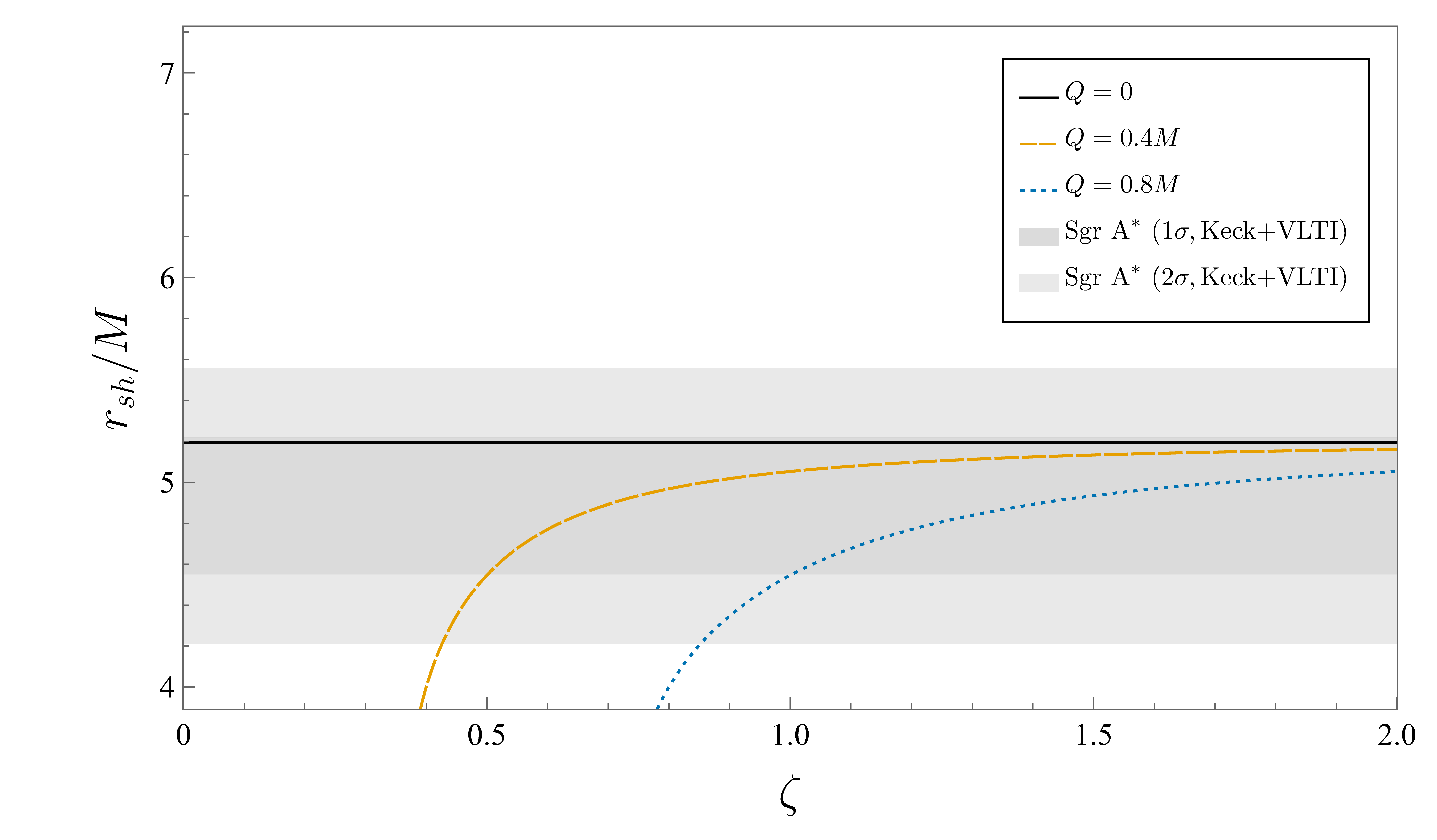}
    \caption{Shadow radius $r_{sh}$ of model III \eqref{eq:model_III} as a function of the correction factor $\zeta$, for three charge values: $Q=0$ (black solid line), $Q=0.4M$ (gold dashed line), $Q=0.8 M$ (dotted blue line); considering an external observer at $r_O =  4.1\times 10^{10}M$. The shaded areas represent the confidence intervals of Sgr A*'s shadow radius at 1$\sigma$ (dark gray) and at 2$\sigma$ (light gray).}
    \label{fig:shadow_IV}
\end{figure}

In Fig.~\ref{fig:shadow_IV}, the shadow radius of the space-time geometry~\eqref{eq:model_III_flat} is plotted as a function of $\zeta$, considering an observer located at $r_O = 4.1\times 10^{10}M$ and $\omega \rightarrow \infty$. We study three separate scenarios of the usual charge: $Q=0$, $Q=0.4M$ and $Q=0.8M$.

Observing Fig.~\ref{fig:shadow_IV}, the suppressive/enhancing role of $\zeta$ is revealed. For $\zeta=1$, the relation between the usual charge and the dressed charge is reduced to $Q^2=\tilde{Q^2}$, meaning the shadow size of each geometry matches that of a RN black hole with the respective charge value. For values of $\zeta < 1$, the correction is enhanced and the shadow size of the geometry remains compatible with the EHT observations for a tighter interval of the charge. Should any constraints of $\zeta$ push it to such values, it therefore places more meaningful constraints in the charge as well. In contrast, for values of $\zeta > 1$, the correction is suppressed and the shadow size of the geometry remains compatible with the EHT observations for a larger interval of the charge, even allowing it to be pushed beyond values for which a RN black hole of comparable charge would predict the formation of a naked singularity. 

To better illustrate the effects of $\zeta$ we generate images of a RN-like black hole within the non-minimally coupled Weyl gravity framework, when surrounded by a thin accretion disk. Specifically, we simulate the optical appearance of charged Weyl black hole as seen by an observer at a face-on inclination (i.e. $\theta_o = 0^{\circ}$) and illuminated by an optically (i.e. fully transparent to its own radiation) and geometrically thin, axisymmetric accretion disk, conveniently modeled by the Gralla-Lupsasca-Marrone (GLM) emission profiles \cite{Gralla:2020srx}. The latter consist of semi-analytic approximations of the simulated time-averaged accretion flow around a Kerr black hole, offering an alternative to the computationally taxing General Relativistic Magneto-Hydrodynamic (GRMHD) simulations. 

The GLM models describe the radial emission profile of an accretion disk based on a Standard Unbound (SU) Johnson distribution function given by
\begin{equation}\label{eq:GLMfunction}
    I(r;\gamma,\mu,\sigma)=\frac{e^{-\frac{1}{2}\left[\gamma+\text{arcsinh}(\frac{r-\mu}{\sigma})\right]^2}}{\sqrt{(r-\mu)^2+\sigma^2}},
\end{equation}
where the $\gamma, \mu$ and $\sigma$ parameters modify the distribution's skewness, peak location and kurtosis, respectively. We assume the disk to be a monochromatic and optically thin emitter, as this simplifies the integration of the relativistic radiative transfer equation along a null geodesic considerably. 

Thus, for a geometrically thin accretion disk, the observed intensity corresponds to a sum of the intensities of each intersection (i.e. every $n$ half-orbit) with the accretion disk
\begin{equation}\label{eq:integratedIntensity}
    I_{o}(r)=\sum_{n}g_{}(r)^2\left.I(r)\right|_{r=r_n}.
\end{equation}
For the purpose of generating images, we restrict out analysis to the $n=2$ photon ring, since higher-order rings enact a negligible contribution to the net luminosity. We employ a Mathematica-based relativistic raytracing code - the Geodesic Rays and Visualization of IntensiTY profiles (GRAVITYp) module~\cite{daSilva:2023jxa}.

\begin{figure*}[t!]
	\subfloat{
		\includegraphics[width=5.30cm,height=4.70cm]{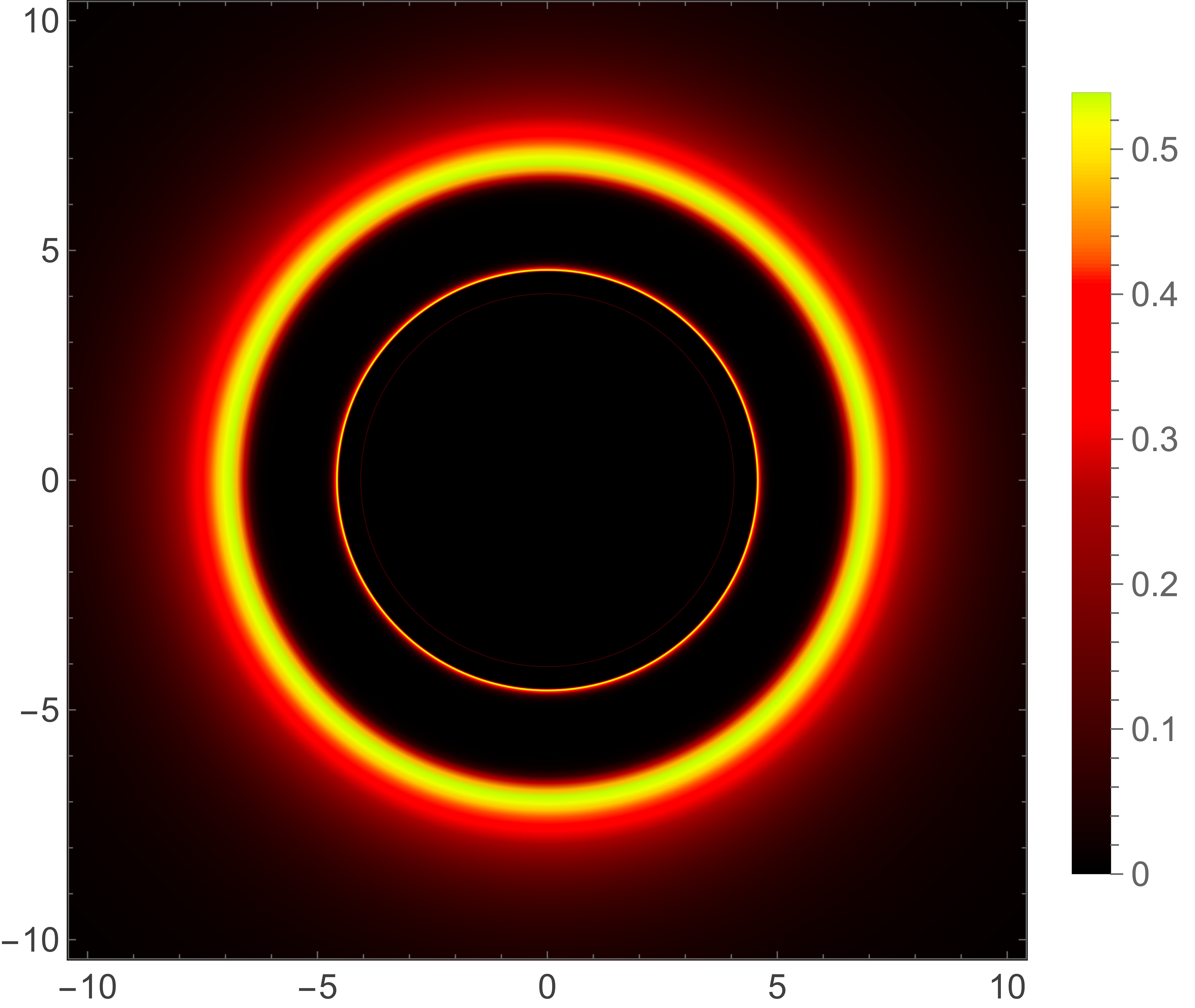}
		\includegraphics[width=5.30cm,height=4.70cm]{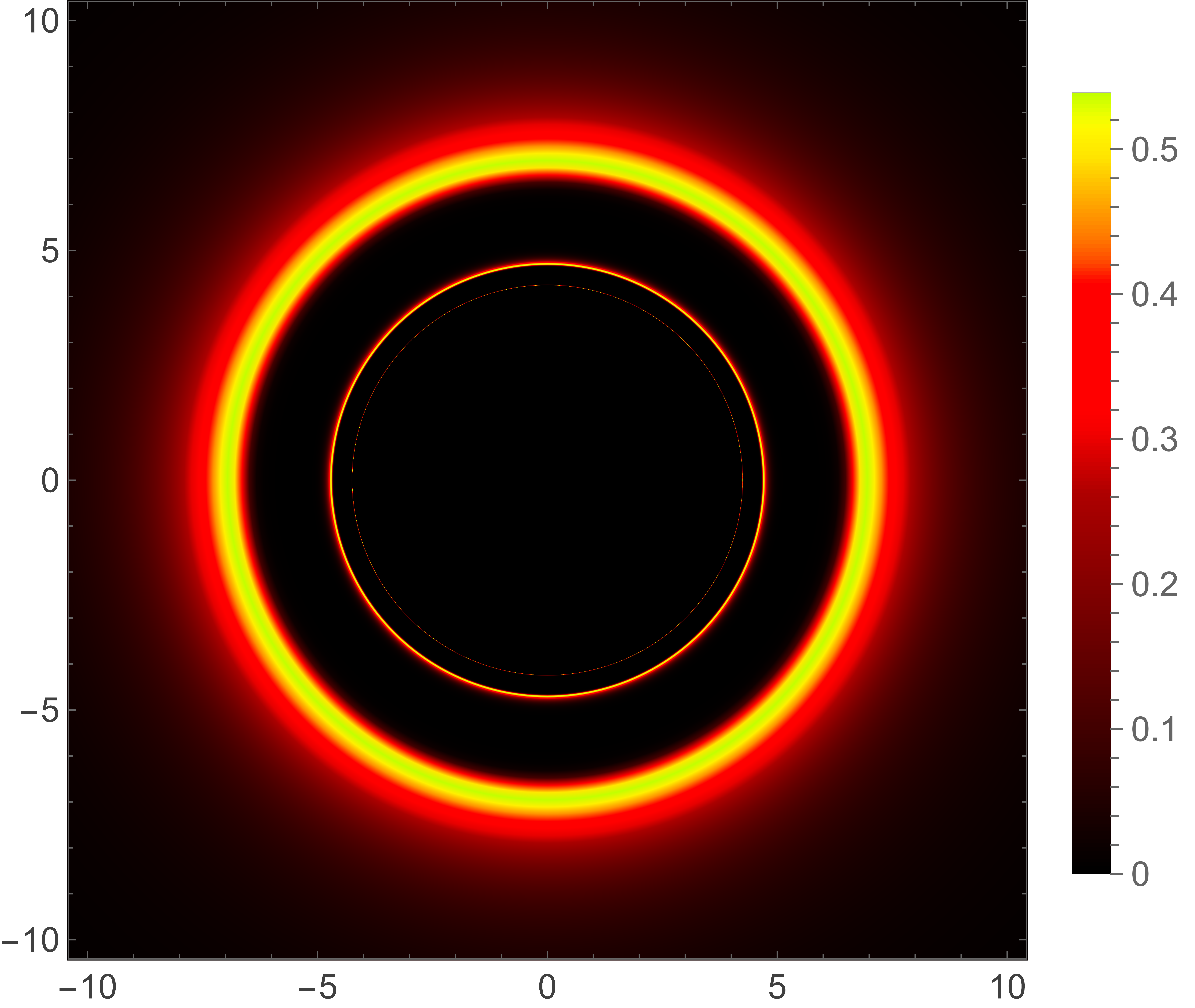}
		\includegraphics[width=5.30cm,height=4.70cm]{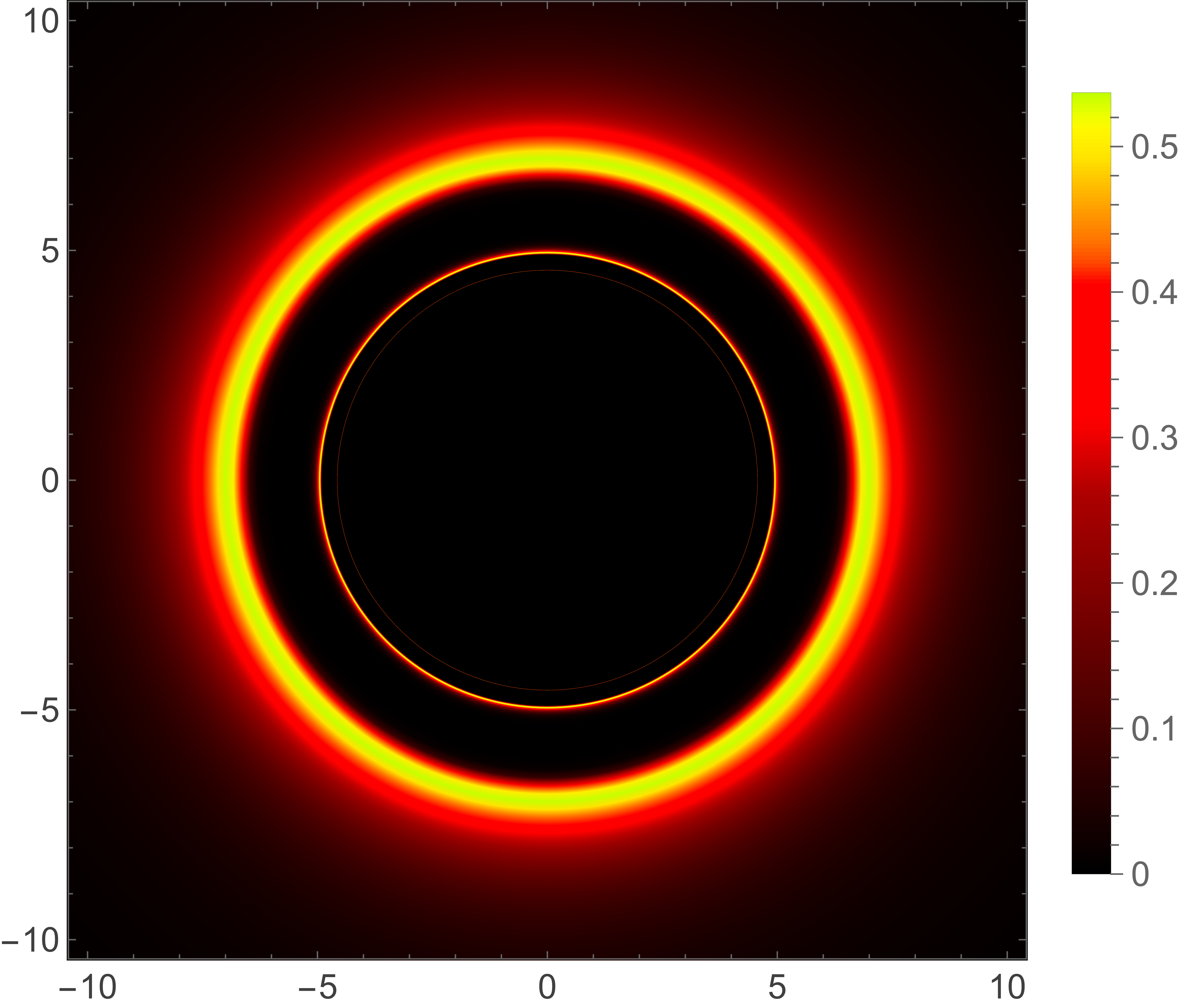}
	}\quad
	\subfloat{
		\includegraphics[width=5.30cm,height=4.70cm]{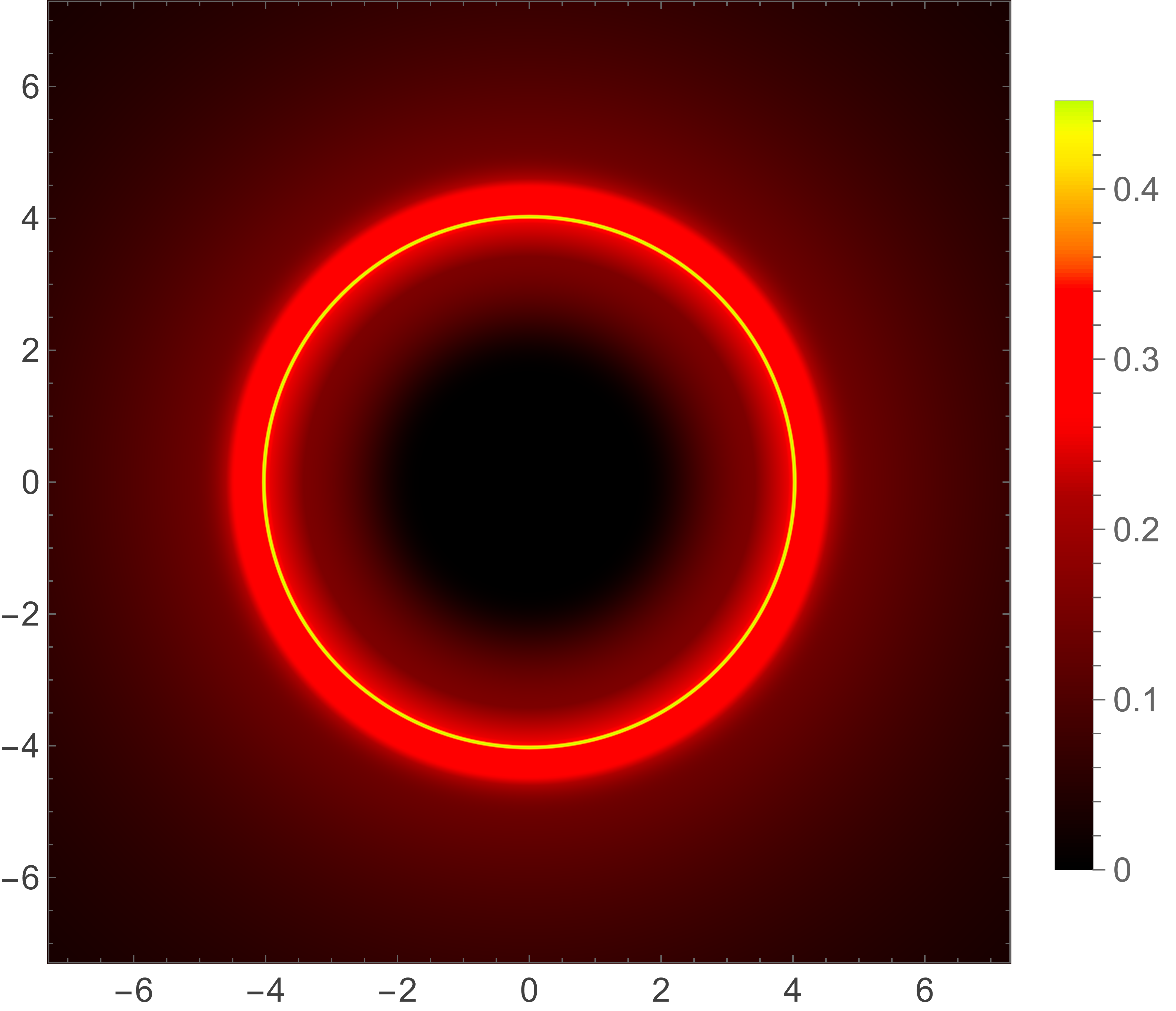}
		\includegraphics[width=5.30cm,height=4.70cm]{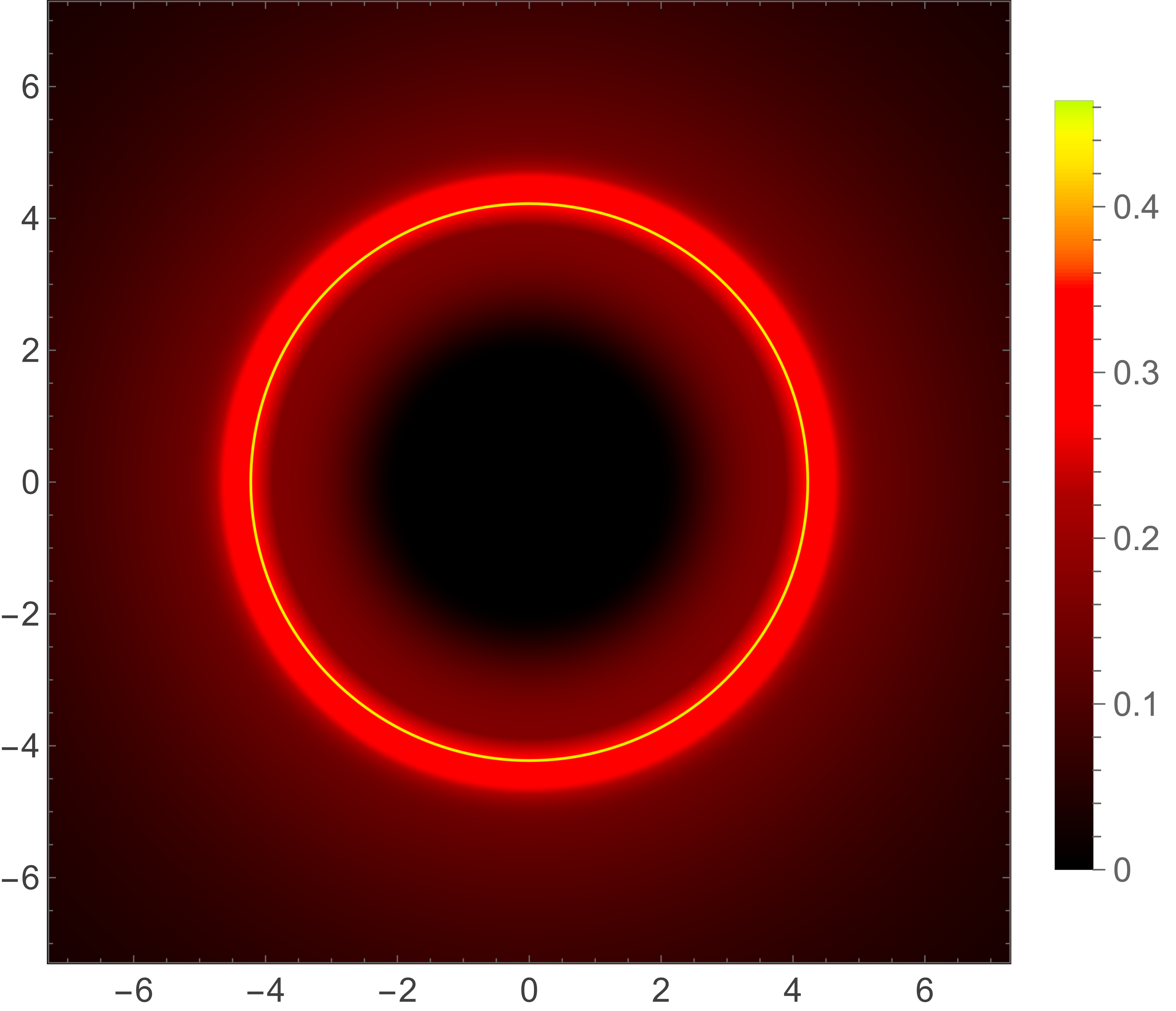}
		\includegraphics[width=5.30cm,height=4.70cm]{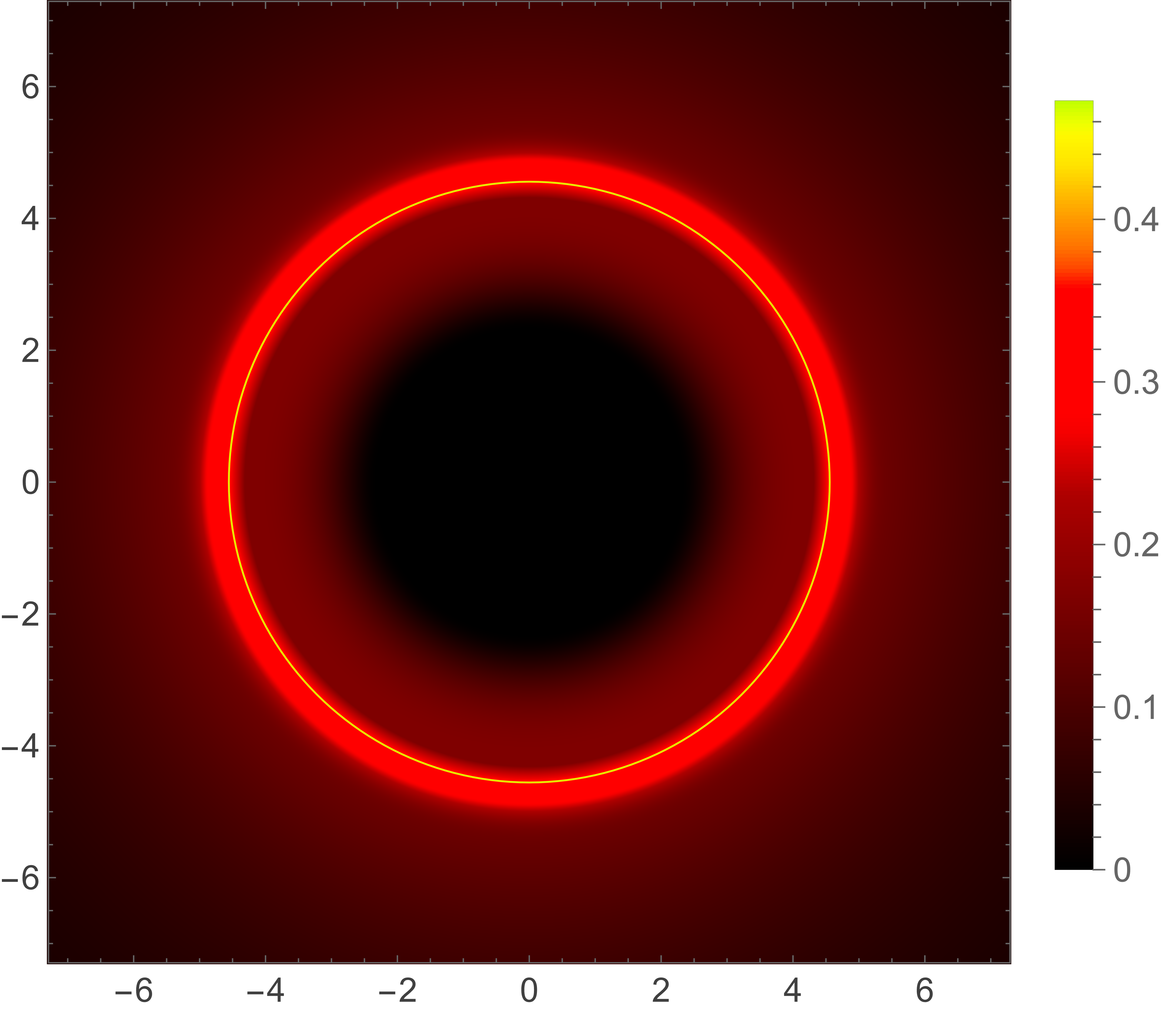}
	}\quad
	\subfloat{
		\includegraphics[width=5.30cm,height=4.70cm]{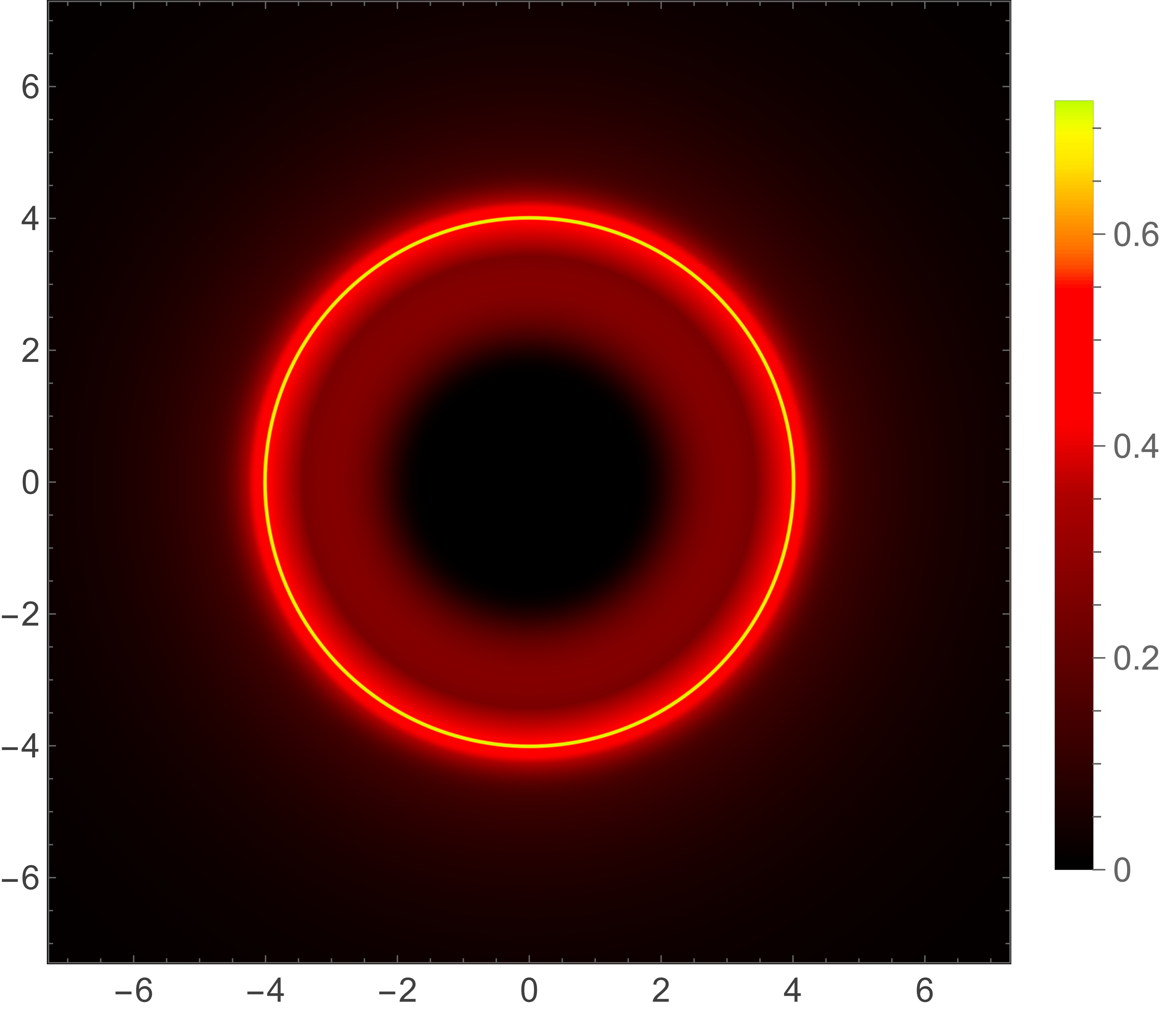}
		\includegraphics[width=5.30cm,height=4.70cm]{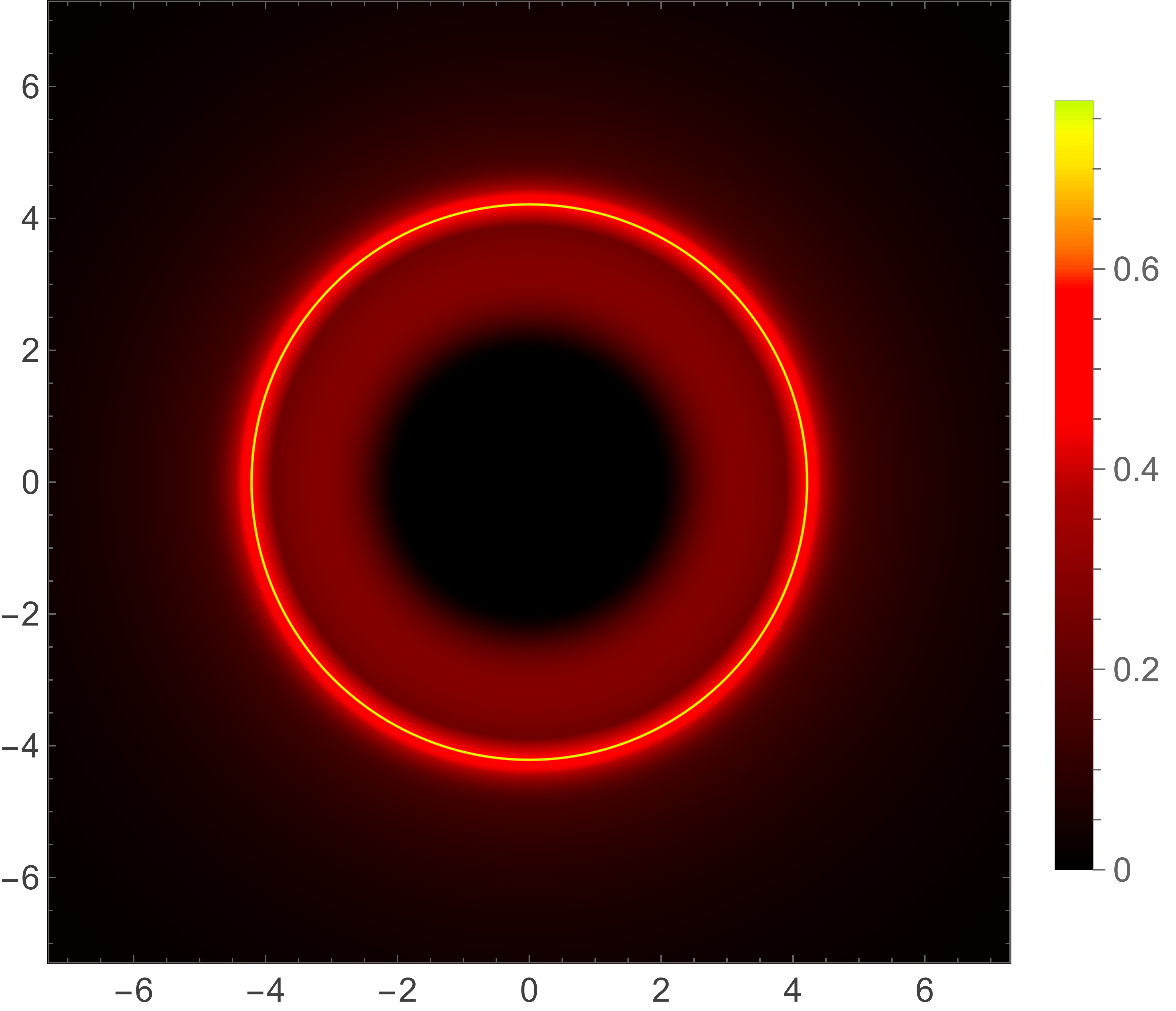}
		\includegraphics[width=5.30cm,height=4.70cm]{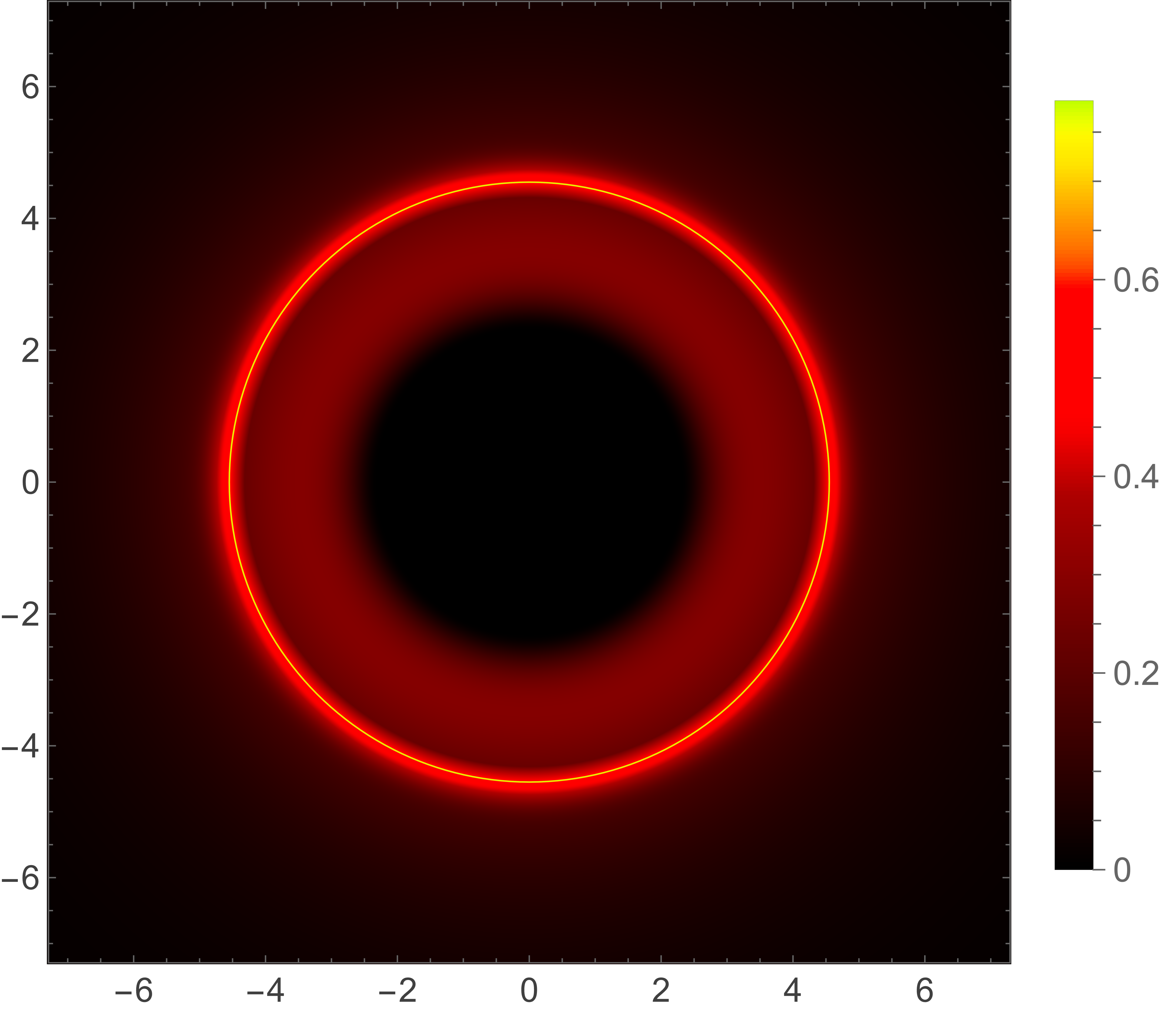}
	}
	\caption{Visual appearance, in the impact parameter space, of the spacetime geometry modeled by equation \eqref{eq:model_III_flat}. The images are organized according to the emission profile: GLM3 (top line), GLM1 (middle line) and GLM2 (bottom line); and the value of the correction factor: $\zeta=0.94$ (left column), $\zeta=1.00$ (middle column) and $\zeta=1.175$ (right column). We consider a Weyl black hole with a charge $Q = 0.94 M$, as seen at a face-on orientation, by an observer located at a distance of $r_O =  4.1\times 10^{10}M$.}
	\label{fig:BHimages}
\end{figure*}

We consider the three models introduced in \cite{Gralla:2020srx}, with the parameters
\begin{eqnarray}
\text{GLM1}&:&   \gamma=-\frac{3}{2} \, , \quad \mu=0 \, , \qquad \  \sigma=\frac{M}{2} \label{eq:GLM1} \ , \\
\text{GLM2}&:&   \gamma=0 \, , \qquad \mu=0 \, , \qquad \  \sigma=\frac{M}{2} \    \label{eq:GLM2}, \\
\text{GLM3}&:& \gamma=-2 \,, \quad \  \mu=\frac{17M}{3} \, , \    \sigma=\frac{M}{4} \label{eq:GLM3}  \ .
\end{eqnarray}
The GLM1 and GLM2 models simulate a disk emission profile that extends all the way to the event horizon, each with a distinct radial emission decay. In contrast, the GLM3 model's emission peak is located near the innermost stable circular orbit radius ($r_{ISCO}$); an emission profile which cleanly separates the individual luminosity contributions of the $n=1$ and $n=2$ rings from the direct emission. In Fig.~\ref{fig:BHimages}, we compare three charged Weyl black holes, with $Q = 0.94 M$ (i.e. the charge value at which a RN black hole saturates the lower $2\sigma$ EHT bound \cite{daSilva:2023jxa}), considering $\zeta \in \{0.94; 1.00; 1.175\}$ respectively. To ensure each setting is compared in identical terms, the observed luminosity is normalized with respect to its total value.

From a qualitative analysis of Fig.~\ref{fig:BHimages} there are standout differences between all of the considered cases. With regard to differences stemming from the accretion model, our results align with others in literature \cite{Gralla:2019xty, Gralla:2020srx}. Each sub-figure line corresponds to a different GLM model, as one can tell from the corresponding direct emission: in the top line the direct emission peaks before the photon ring emission, as intended by this particular profile; in the middle line of images, the direct emission appears as a smooth dark red background covering most of the images' panels; the direct emission in the bottom line has a comparably a sharper decay. Conversely, for the GLM1 and GLM2 models, the location of the ring emission remains unchanged, particularly with regard to the $n=2$. This is somewhat true for the GLM3 model, though the $n=2$ ring emission is quite faint.

To compare the models with respect to their value of the correction factor we focus on the images on a column-by-column basis, where the left-most images correspond to a geometry with a smaller value of $\zeta$ and the right-most images correspond to a geometry with a larger value of $\zeta$. Decreasing the value of $\zeta$ enhances the effect of the charge, resulting in an overall lower shadow radius. This is confirmed by smaller size of the photon rings, which asymptote to the edge of the black hole shadow, in the left column images. On the other hand, a higher value of $\zeta$ suppresses the effect of the charge, resulting in a larger sized shadow, and therefore larger photon rings. It is worth noting that not only the sizes differ, but also the width of the $n=1$ and $n=2$ rings, as can be seen by the thickness differences of the bright red and bright yellow rings in each column. 

The GLM3 model allows us to further appreciate such differences without contamination of the direct emission. The choice of $Q$ and $\zeta$ values are not arbitrary. In fact, the geometries represented in the middle column, for which $\zeta=1.00$ and $Q = 0.94 M$, coincide with a RN black hole with the maximum charge value before saturating the lower $2\sigma$ EHT shadow bound. These results, along with those of Fig.~\ref{fig:shadow_IV}, demonstrate therefore that depending on the value of $\zeta$ the actual charge can be higher or lower before exceeding the compatibility bounds imposed by the EHT observations. At $\zeta=0.94$ and $Q=0.94 M$, the resulting dressed charge is $\Tilde{Q}=1.00 M$ which is the critical charge value beyond which the geometry becomes a RN naked singularity. For $\zeta=1.175$ and $Q=0.94 M$, the resulting dressed charge is $\Tilde{Q}=0.8 M$. In other words, the presence of the $\zeta$ factor allows the charge of a Weyl black hole to exceed that of RN black hole before its shadow size becomes incompatible with current observational data. As long as the effective charge remains below $\Tilde{Q}=0.94 M$, the latter remain true.

\section{Summary and Conclusion}\label{sec:conc}

This work has explored black hole shadows in the context of nonminimally coupled Weyl connection gravity, a class of alternative theories in which the spacetime geometry is characterized by both the metric tensor and a Weyl vector field. A key aspect of this framework is the nonminimal coupling between matter and geometry, together with the presence of a non-dynamical Weyl vector encoding spacetime non-metricity. Owing to this structure, the resulting field equations remain second order in derivatives, in contrast with many higher-curvature extensions of general relativity (GR), thereby avoiding the typical higher-derivative instabilities that often arise beyond GR.

In this setting, Schwarzschild- and Reissner--Nordstr\"{o}m-like black hole solutions naturally emerge. These geometries deviate from their GR counterparts through additional contributions governed by the Weyl integration constant associated with the vector-field solution. Under appropriate conditions, such corrections preserve the overall static and spherically symmetric character of the spacetime, while introducing modifications that directly encode the effects of non-metricity.

Motivated by horizon-scale observations, we have computed the shadows cast by these black hole solutions and used them to constrain the Weyl constant. Fixing the observer’s distance to $r_O = 4.1\times 10^{10}M$ in geometrized units, corresponding to the average of the Keck and VLTI measurements of the distance to Sgr A*, we imposed compatibility with the EHT shadow size at the $2\sigma$ confidence level. Our analysis yields the bounds $\omega \gtrsim 10^{11.7}M$ for model I, $\omega \gtrsim 10^{10.5}M$ for model II, and a starting bound of $\omega \sim 10^{11.7}M$ for model III, which becomes more relaxed as the dressed charge increases. These results show that current black hole imaging data already place meaningful constraints on departures from metric compatibility in Weyl-based theories.

It is important to emphasize, however, that our approach relies on fractional deviations calibrated within asymptotically flat, GR-based scenarios. Since the EHT analysis assumes backgrounds close to Kerr geometry, applying those constraints to Weyl spacetimes may introduce a residual theoretical bias. A fully self-consistent assessment would require dedicated ray-tracing simulations and image reconstructions performed directly within the Weyl framework, including a re-evaluation of the calibration factors entering the shadow inference procedure.

Furthermore, the constraints derived here can be interpreted in the broader context of testing the Schiff conjecture, which states that Local Position Invariance, the Equivalence Principle, and Local Lorentz Invariance are interconnected \cite{Schiff:1960ggq,Tasson:2016xib}. Non‑metricity in Weyl gravity is intimately related to violations of the Einstein Equivalence Principle, e.g., through the existence of a preferred frame or a fifth force. Our bounds on the Weyl constant $\omega$ imply that any such violation must be extremely small on galactic scales, providing a complementary observational constraint to those obtained from solar system or laboratory experiments. Similar tests of Lorentz violation have been performed in bumblebee gravity using black hole shadows \cite{Capozziello:2023rfv,Capozziello:2023tbo,Casana:2017jkc}, and our results extend this line of inquiry to Weyl‑based theories.

Looking ahead, several directions naturally follow from the present study. First, extending the analysis to rotating solutions would be essential, as spin-induced effects could amplify deviations from GR and provide complementary observational signatures. Second, incorporating additional observables, such as photon ring substructure, lensing signatures, or quasinormal mode spectra, could strengthen the phenomenological constraints on non-metricity. Third, revisiting black hole properties derived in Boyer-Lindquist coordinates and generalizing them consistently within Weyl geometries would clarify the robustness of previously established results. Ultimately, combining shadow observations with independent probes, such as gravitational-wave measurements, may offer a comprehensive strategy to test the role of non-metricity and further assess the viability of Weyl connection gravity as an extension of GR.

	\acknowledgments{
		ML acknowledges support from Fundo Regional da Ciência e Tecnologia and Azores Government through the Fellowship M3.1.a/F/031/2022, and from FCT/Portugal through CAMGSD, IST-ID, projects UID/04459/2025 and the H2020-MSCA-2022-SE project EinsteinWaves, GA No.101131233.
		FSNL acknowledges support from the Funda\c{c}\~{a}o para a Ci\^{e}ncia e a Tecnologia (FCT) Scientific Employment Stimulus contract with reference CEECINST/00032/2018, and funding through the grants UID/04434/2025, and PTDC/FIS-AST/0054/2021.
		FSNL acknowledges support from the Funda\c{c}\~{a}o para a Ci\^{e}ncia e a Tecnologia (FCT) Scientific Employment Stimulus contract with reference CEECINST/00032/2018, and funding through the grants UID/04434/2025, and PTDC/FIS-AST/0054/2021.}
	

	
\end{document}